\documentstyle[12pt]{article}
\begin{document}

\baselineskip=30pt

\begin{titlepage}

\vskip1in

\begin{center}
\LARGE{\bf Mass Scales and Their Relations in Symmetric Quantum Field Theory.}

\vskip1.0cm

\Large{ S. R. Gobira$^{1}$ \ and  M. C. Nemes$^{2}$}\\ 
\end{center} 

\vskip1.0cm
\begin{center}
$(1)$UFT - Universidade Federal do Tocantins\\
N\'{u}cleo de F\'{i}sica\\
CEP-77020-210, Palmas - TO - Brazil\\
e-mail gobira@uft.edu.br\\
\vskip1.0cm
$(2)$UFMG - Universidade Federal de Minas Gerais\\
Departamento de F\'{i}sica - ICEx\\
C.P.702, CEP 30.161-970, Belo Horizonte - MG - Brazil\\
e-mail carolina@fisica.ufmg.br\\
\vskip1.0cm

\end{center}
\vskip0.5cm

\newpage
\begin{abstract}
\noindent
We illustrate the importance of mass scales and their relation in the specific case of the linear sigma model within the context of its one loop Ward identities. In the calculation it becomes apparent the delicate and essential connection between divergent and finite parts of amplitudes. The examples show how to use mass scales identities which are absolutely necessary to manipulate graphs involving several masses. Furthermore, in the context of the Implicitly Regularization, finite(physical) and divergent (counterterms) parts of the amplitude can and must be written in terms of a single scale which is the renormalization group scale. This facilitates, e.g., obtaining symmetric counterterms and immediately lead to the proper definition of Renormalization Group Constants.

\end{abstract}
\end{titlepage}

\section{\protect\bigskip Introduction}

A field theory predictivity, or the ability to obtain results valid to all
orders of perturbation theory, relies on logical conditions: the
renormalization program has to be a systematic and unambiguously fixed
algorithm that satisfies the fundamental properties of locality and
causality: it should correspond to the addition of local counterterms to the
Lagrangian density. In general, any renormalization procedure involves two
steps \cite{bonn}: 1) A choice of Regularization followed by a subtraction
procedure. 2) A set of renormalization conditions in order to define the
parameters of theory at a given scale.

This scale acquires a very crucial role. From \ the most \ naive point of
view it is an arbitrariness coming from the fact that the separation of a
divergent amplitude in a finite plus divergent part is defined up to a
constant. In several approaches in the literature it appears in different
ways, for example in Dimensional Regularization \cite{RD} it appears for
dimensional reasons. In Differential Renormalization \cite{Dif.R} it appears
as an integration constant. In our scheme it appears in the rather subtle
way, as we will see.

The purpose of this rather technical work is to perform an analytical
evaluation of all amplitudes at one loop level which are necessary for an 
{\em explicit }verification of the Ward identities of the chiral linear
sigma model with fermions. In this work we illustrate the importance of mass
scales and their relation in the specific case of this model within the
context of its one loop Ward identities. Besides, we use the unrenormalized
amplitude to illustrate the communication between finite(physical) and
divergent (counterterms) parts of the amplitude in order to check the Ward
identities. The examples show how to use mass scales identities which are
absolutely necessary to manipulate graphs involving several masses.
Furthermore, in the context of the Implicitly Regularization \cite{batis}
\cite{irt}\cite{irt2}, finite(physical) and divergent (counterterms) parts
of the amplitude can and must be written in terms of a single scale which is
the renormalization group scale. This facilitates, e.g., obtaining symmetric
counterterms and immediately leads to the proper definition of
Renormalization Group Constants.

Our technique to handle such amplitudes is to assume only implicitly the
presence of a regulator in the integrals and algebraically manipulate the
integrand until such a separation is achieved. The divergent parts are left
in the form of integrals which enable us to recover the result of any
regularization prescription.

The reason we have chosen the chiral model is related to the presence of the 
$\gamma ^{5}$ matrix, and to show it can be handled in 4-dimensions without
difficulties. The presence of three different masses in this model help us
illustrate the scale change mechanism in the context of Ward identities.

This work is organized as follows: in the section 2 we present the model and
the relevant Ward-Takahaski identities . In section 3 we verify that chiral
symmetry helps ''taming'' the divergent content of this model. In the
section 4 we briefly recall the Implicit Regularization Prescription. In the
section 5 we show the mechanism of using the relations between mass scales.
Final comments can be found in section 6.

\section{The linear sigma model}

The linear sigma model has a renormalizable Lagrangian constructed by J.
Schwinger, J.C \cite{sig}. Polkingorne \cite{sig3}, M. Gell-Mann and M.Levy 
\cite{sig2}. In this model the fields are: \ $N_{o}(p,n)$ \ the nucleon
isodublect (fermions), $\stackrel{\rightarrow }{\pi _{o}}(\pi _{1},\pi
_{2},\pi _{3})$ pion isotriplect (pseudoscalar), \ $\sigma _{o}$ sigma
(isoscalar). The chiral symmetric Lagrangian is 
\begin{eqnarray}
{\cal L}_{o} &=&i\bar{N}_{o}\not{\partial}N_{o}+\frac{1}{2}\left[ (\partial
_{\mu }\sigma _{o})^{2}+(\partial _{\mu }\vec{\pi}_{o})^{2}\right] \\
&&-\frac{\mu _{o}^{2}}{2}(\sigma _{o}^{2}+\vec{\pi}_{o}^{2})-\frac{\lambda
_{o}}{4}(\sigma _{o}^{2}+\vec{\pi}_{o}^{2})^{2}  \nonumber \\
&&-G_{o}\bar{N}_{o}(\sigma _{o}+i\gamma ^{5}\vec{\tau}.\vec{\pi}_{o})N_{o} 
\nonumber
\end{eqnarray}
The chiral breaking term is usually

\begin{equation}
{\cal L}_{q}=c_{o}\sigma _{o}
\end{equation}
where\ $c_{o}$ \ is a parameter.

The explicit chiral symmetry breaking terms, as is well known, will give
rise to a nonvanishing vacuum expectation value of the sigma field $<\sigma
_{o}>=v_{o}$. We therefore perform a shift in order to have the Lagrangian
in terms of fields with zero expectation value as below. 
\begin{equation}
s_{o}=\sigma _{o}-v_{o}  \label{17}
\end{equation}

The redefined Lagrangian reads 
\begin{equation}
{\cal L}_{T}={\cal L}_{F}+{\cal L}_{I}  \label{18}
\end{equation}
where 
\begin{eqnarray}
{\cal L}_{F} &=&\bar{N}_{o}(i\not{\partial}-G_{o}v_{o})N_{o}  \nonumber \\
&&+1/2[(\partial _{\mu }s_{o})^{2}-(\mu _{o}^{2}+3\lambda
_{o}v_{o}^{2})s_{o}^{2}]  \nonumber \\
&&+1/2[(\partial _{\mu }\vec{\pi}_{o})^{2}-(\mu _{o}^{2}+\lambda
_{o}v_{o}^{2})\vec{\pi}_{o}^{2}]  \label{19}
\end{eqnarray}
and 
\begin{eqnarray}
{\cal L}_{I} &=&-G_{o}\bar{N}_{o}(s_{o}+i\gamma ^{5}\vec{\tau}.\vec{\pi}%
_{o})N_{o}  \nonumber \\
&&-\lambda _{o}/4(s_{o}^{2}+\vec{\pi}_{o}^{2})^{2}  \nonumber \\
&&-\lambda _{o}v_{o}(s_{o}^{2}+\vec{\pi}_{o}^{2})s_{o}  \label{20}
\end{eqnarray}
We can now read off the nucleon, pion and sigma meson masses 
\begin{equation}
M_{N}=G_{o}v_{o}  \label{21}
\end{equation}
\begin{equation}
M_{\pi }^{2}=\mu _{o}^{2}+\lambda _{o}v_{o}^{2}  \label{22}
\end{equation}
\begin{equation}
M_{\sigma }^{2}=\mu _{o}^{2}+3\lambda _{o}v_{o}^{2}  \label{23}
\end{equation}
We also define, for purposes of Ward Identities, seven vertices functions,
with mesons having zero momentum and the fermions on their mass shell 
\begin{equation}
V_{N\pi N},V_{N\sigma N},V_{\pi ^{4}},V_{\sigma ^{4}},V_{\pi ^{2}\sigma
^{2}},V_{\pi ^{2}\sigma },V_{\sigma ^{3}}
\end{equation}
The indices indicate the interaction involved. The set of Feynman rules we
have used are those of Ref. \cite{BT} with $F=<\sigma >$ , $\sigma $ being
the renormalized field. The renormalized quantities are introduced as
follows, 
\begin{eqnarray}
N_{o} &=&\sqrt{Z_{F}}N  \nonumber \\
(s_{o},v_{o},\vec{\pi}_{o}) &=&\sqrt{Z_{B}}(s,v,\vec{\pi})  \nonumber \\
\mu _{o}^{2} &=&\frac{1}{Z_{B}}(\mu ^{2}+\delta \mu ^{2})  \nonumber \\
G_{o} &=&Z_{g}/(Z_{F}\sqrt{Z_{B}})G  \nonumber \\
\lambda _{o} &=&(Z_{\lambda }/Z_{B}^{2})\lambda 
\end{eqnarray}
which are enough to render the model finite.

\section{A Little Help from Chiral Symmetry}

Now we will verify that chiral symmetry helps ''taming'' the divergent
content of the model. The Ward Identities relate three point functions
(logarithmically divergent) with two points functions (quadratically and
linearly divergent). In a general way this fact allows to establish
algebraic relations among the subtraction constants of the divergent Green
functions of a theory and, therefore, it is a useful tool in the
renormalization procedure. Nevertheless Ward Identities can be violated by
some used regularization scheme \cite{bjp1} like gauge symmetry in the
Quantum Electrodynamics. For the linear sigma model chiral symmetry it is
really regularization independent. In an explicit calculation of one loop
amplitudes we verify that all quadratically and linearly divergent integrals
cancel in the verification of the relative Ward Identities. This
cancellation occurs at the level of the evaluation of the Dirac trace and
therefore no specific regularization method is required.

We will use the mass parameters $\mu $, $M$ and $m$ corresponding
respectively to the free fields of the pion, sigma and nucleon. The first
Ward identity that we consider is 
\begin{equation}
-F[V_{\pi ^{2}\sigma }(p,0)]=D_{\sigma }^{-1}(p^{2})-D_{\pi }^{-1}(p^{2})
\label{iw1}
\end{equation}
where $D_{\sigma }(p^{2})$ and $D_{\pi }(p^{2})$ are the sigma field
propagator and pion field respectively, $V_{\pi ^{2}\sigma }(p,0)$ is the
three point function with a pion leg at zero external momentum. The second
Ward identity can be written as 
\begin{equation}
iF[V_{N\pi N}(p,0)]=\frac{1}{2}\left\{ \tau _{a}\gamma _{5},S^{-1}(\not%
{p})\right\} .  \label{iw2}
\end{equation}
In the expression above $\{\}$ means anticomutator, $S(\not{p})$ is the
nucleon propagator and $V_{N\pi N}(p,0)$ is another three point function.
Only two point functions have quadratic and linear divergences. We therefore
conclude that the above two Ward Identities will be enough to verify the one
loop cancellation of the integrals containing quadratic and linear
divergences.

Let us first analyze the\ identity in equation (\ref{iw1}). The inverse of
the pion propagator is given by 
\begin{equation}
D_{\pi }^{-1}(p^{2})=p^{2}-\mu ^{2}-\Sigma _{R}^{\pi }(p^{2})
\end{equation}
where 
\begin{equation}
\Sigma _{R}^{\pi }(p^{2})=\Sigma _{CT}^{\pi }(p^{2})+\Sigma ^{\pi }(p^{2})
\end{equation}
the indices indicate renormalized amplitudes R and counterterms are
indicated by CT. Since the contribution comes from only two diagrams, we
write 
\begin{equation}
\Sigma ^{\pi }(p^{2})=\Sigma _{1}^{\pi }(p^{2})+\Sigma _{2}^{\pi }(p^{2})
\end{equation}
where the indices 1 and 2 indicate the contributions of each diagram. The
explicit perturbative expressions for this contributions are given in terms
of Feynman amplitudes and can be written as 
\begin{equation}
i\Sigma _{1}^{\pi }(p^{2})=8G^{2}[p_{\mu }I_{lin}^{\mu
}(p^{2},m^{2})-I_{quad}(p^{2},m^{2})]
\end{equation}
and 
\begin{equation}
i\Sigma _{2}^{\pi }(p^{2})=4F^{2}\lambda ^{2}I_{\log }(p^{2},\mu ^{2},M^{2})
\end{equation}
where 
\begin{equation}
I_{lin}^{\mu }(p^{2},m^{2})=\int_{\Lambda }\frac{d^{4}k}{(2\pi )^{4}}\frac{%
k^{\mu }}{(k^{2}-m^{2})[(p-k)^{2}-m^{2}]},
\end{equation}
\begin{equation}
I_{quad}(p^{2},m^{2})=\int_{\Lambda }\frac{d^{4}k}{(2\pi )^{4}}\frac{1}{%
[(p-k)^{2}-m^{2}]}
\end{equation}
and 
\begin{equation}
I_{\log }(p^{2},\mu ^{2},M^{2})=\int_{\Lambda }\frac{d^{4}k}{(2\pi )^{4}}%
\frac{1}{(k^{2}-M^{2})[(p-k)^{2}-\mu ^{2}]}
\end{equation}
The index $\Lambda $ in the integral means only an implicit regularization
and a specific regulator needs never be used.

We also have the inverse sigma propagator at one loop order. 
\begin{equation}
D_{\sigma }^{-1}(p^{2})=p^{2}-M^{2}-\Sigma _{R}^{\sigma }(p^{2})
\end{equation}
In the same way as before 
\begin{equation}
\Sigma _{R}^{\sigma }(p^{2})=\Sigma _{CT}^{\sigma }(p^{2})+\Sigma ^{\sigma
}(p^{2})
\end{equation}
and since the contribution comes from three diagrams, we write 
\begin{equation}
\Sigma ^{\sigma }(p^{2})=\Sigma _{1}^{\sigma }(p^{2})+\Sigma _{2}^{\sigma
}(p^{2})+\Sigma _{3}^{\sigma }(p^{2}).
\end{equation}
Here 
\begin{equation}
i\Sigma _{1}^{\sigma }(p^{2})=8G^{2}[p_{\mu }I_{lin}^{\mu
}(p^{2},m^{2})-I_{quad}(p^{2},m^{2})-2m^{2}I_{\log }(p^{2},m^{2},m^{2})],
\end{equation}
\begin{equation}
i\Sigma _{2}^{\sigma }(p^{2})=18F^{2}\lambda ^{2}I_{\log }(p^{2},M^{2},M^{2})
\end{equation}
and 
\begin{equation}
i\Sigma _{3}^{\sigma }(p^{2})=6F^{2}\lambda ^{2}I_{\log }(p^{2},\mu ^{2},\mu
^{2})
\end{equation}
The cancellation of the quadratically and linearly divergent contributions
can be seen in only the terms $\Sigma _{1}^{\pi }(p^{2})$ and $\Sigma
_{1}^{\sigma }(p^{2})$. Since the integrals $I_{quad}(p^{2},m^{2})$ and $%
I_{lin}^{\mu }(p^{2},m^{2})$ are the same for both contributions ( $\Sigma
_{1}^{\pi }(p^{2})$ and $\Sigma _{1}^{\sigma }(p^{2})$.) and since the Ward
Identity is given by the difference between them, the cancellation is
obvious and no specific regularization method is required.

Let us to analyze the second Ward identity (\ref{iw2}). In this case we must
consider the inverse of the nucleon propagator including its one loop
correction 
\begin{equation}
S_{F}^{-1}(\not{p})=\not{p}-m+\Sigma _{R}^{N}(\not{p})
\end{equation}
(we used the notation $\not{p}=\gamma ^{\mu }p_{\mu }$).In an analogous
fashion as in the preceding case, we have 
\begin{equation}
\Sigma _{R}^{N}(\not{p})=\Sigma _{CT}^{N}(\not{p})+\Sigma ^{N}(\not{p})
\end{equation}
and as the contribution comes from two diagrams, we get 
\begin{equation}
\Sigma ^{N}(\not{p})=\Sigma _{1}^{N}(\not{p})+\Sigma _{2}^{N}(\not{p})
\end{equation}
The explicit perturbative expressions for this contributions are given in
terms of Feynman amplitudes 
\begin{equation}
i\Sigma _{1}^{N}(\not{p})=G^{2}\int_{\Lambda }\frac{d^{4}k}{(2\pi )^{4}}%
\frac{\not{k}+m}{(k^{2}-m^{2})[(p-k)^{2}-M^{2}]}
\end{equation}
and 
\begin{equation}
i\Sigma _{2}^{N}(\not{p})=3G^{2}\int_{\Lambda }\frac{d^{4}k}{(2\pi )^{4}}%
\frac{-\not{k}+m}{(k^{2}-m^{2})[(p-k)^{2}-\mu ^{2}]}
\end{equation}
Since $\{\gamma ^{\mu },\gamma ^{5}\}=0$, by the Ward identity (\ref{iw2})
one can see the cancellation of the linearly divergent contributions and as
before no specific regularization method is required.

\section{The Implicit Regularization}

Now a word about the Implicit Regularization Technique which we are using
are in order. A simple example of its working procedure can be found in
several references \cite{batis}\cite{irt}\cite{irt2}, and we give here a
simple illustration. In order to illustrate the procedure, consider the
following divergent amplitude, typical of one loop order: 
\begin{equation}
\int_{\Lambda }\frac{d^{4}k}{(2\pi )^{4}}\frac{1}{%
[(k+p)^{2}-m^{2}](k^{2}-m^{2})}\cdot   \label{1}
\end{equation}
The symbol $\Lambda $ under the integral sign presupposes, as discussed, an
implicit regularization. Now, in order to separate the logarithmic
divergence from the finite part, we use the following identity in the factor
involving the external momentum $p$: 
\begin{equation}
\frac{1}{[(k+p)^{2}-m^{2}]}=\sum_{j=0}^{N}\frac{\left( -1\right) ^{j}\left(
p^{2}+2p\cdot k\right) ^{j}}{\left( k^{2}-m^{2}\right) ^{j+1}}+\frac{\left(
-1\right) ^{N+1}\left( p^{2}+2p\cdot k\right) ^{N+1}}{\left(
k^{2}-m^{2}\right) ^{N+1}[\left( k+p\right) ^{2}-m^{2}]}\cdot   \label{2}
\end{equation}
In the above expression $N$ is chosen so that the last term is finite under
integration over $k$. Notice also that in the first term in equation (\ref{2}%
), the external momentum appears only in the numerator and thus after
integration it can yield at most polynomials in $p$ multiplied by
divergences. For our present example we need $N=0$, since we are dealing
with a logarithmic divergence. We can rewrite (\ref{1}) using (\ref{2}) as 
\begin{equation}
I=\int_{\Lambda }\frac{d^{4}k}{(2\pi )^{4}}\frac{1}{(k^{2}-m^{2})^{2}}-\int 
\frac{d^{4}k}{(2\pi )^{4}}\frac{p^{2}+2p\cdot k}{%
[(k+p)^{2}-m^{2}](k^{2}-m^{2})^{2}}\,\cdot   \label{3}
\end{equation}
Now only the first of these two integrals is divergent. The others can be
easily integrated out to yield 
\begin{equation}
I=I_{log}(m^{2})-\frac{i}{(4\pi )^{2}}Z_{0}(m^{2},p^{2})  \label{4}
\end{equation}
where 
\begin{equation}
I_{log}(m^{2})=\int_{\Lambda }\frac{d^{4}k}{(2\pi )^{4}}\frac{1}{%
(k^{2}-m^{2})^{2}}  \label{5}
\end{equation}
and 
\begin{equation}
Z_{0}(m^{2},p^{2})=\int_{0}^{1}dz\,\ln \Big(\frac{p^{2}z(1-z)-m^{2}}{-m^{2}}%
\Big).\cdot   \label{6}
\end{equation}
Note that, since no explicit form for the regulator has been used, one can
make immediate contact with other regularizations. Details of calculations
of several one loop amplitudes and their associated Ward identities by using
this method can be found in \cite{irt}.

\section{ Scale Change Mechanism}

In this section we will present the mechanism of using the relations between
mass scales skillfully. We will see how it is possible in this scheme to
change scales in the Feynman amplitudes and how this change is intimately
connected to the finite parts of these same contributions. In fact, this
mechanism is quite general and comes stems from the freedom we have when
separating , by means of an identity, a finite from a divergent contribution
in a Feynman amplitude. This becomes specially useful and adequate in
theories involving different masses, as is the case of the linear sigma
model or the SU(3) version of the Nambu and Jona Lasinio model. As we will
see, this mechanism is rather essential in order to verify the Ward
identities by using the complete amplitude, i.e., divergent plus finite
contributions.

Let us begin by using the two definitions already used in the IRT \cite
{batis}. We start with a definition for basic divergent object. The log
divergent quantity : 
\begin{equation}
I_{\log }(\xi _{1}^{2})=\int_{\Lambda }\frac{d^{4}k}{(2\pi )^{4}}\frac{1}{%
(k^{2}-\xi _{1}^{2})^{2}}  \label{93}
\end{equation}
as explained in the previous section , the index $\Lambda $ in the integral
means only an implicit regularization and $\xi _{1}$ is any mass.
Alternatively we can use the identity 
\begin{equation}
\frac{1}{(k^{2}-\xi _{1}^{2})}=\frac{1}{(k^{2}-\xi _{2}^{2})}-\frac{(\xi
_{2}^{2}-\xi _{1}^{2})}{(k^{2}-\xi _{2}^{2})(k^{2}-\xi _{1}^{2})}
\label{ide}
\end{equation}
and integrate without restrictions the finite integrals in order to obtain
the identity 
\begin{equation}
I_{\log }(\xi _{1}^{2})=I_{\log }(\xi _{2}^{2})-\frac{i}{(4\pi )^{2}}\ln
\left( \frac{\xi _{1}^{2}}{\xi _{2}^{2}}\right) .  \label{red}
\end{equation}
Notice that in a theory without mass an infra-red regulator can always be
introduced using the identity (\ref{ide}) with $\xi _{1}=0$ . The function
which identifies all one loop finite parts can always be written as 
\begin{equation}
Z_{0}(\xi _{1}^{2},\xi _{2}^{2},p^{2};\xi _{2}^{2})=\int_{0}^{1}dx\ln \left( 
\frac{p^{2}x(1-x)+x(\xi _{1}^{2}-\xi _{2}^{2})-\xi _{1}^{2}}{(-\xi _{2}^{2})}%
\right) ,
\end{equation}
where the function $Z_{0}(\xi _{1}^{2},\xi _{2}^{2},p^{2};\xi _{2}^{2})$ is
defined having  $\xi _{2}$ as mass scale and then we can use the identity 
\begin{equation}
Z_{0}(\xi _{1}^{2},\xi _{2}^{2},p^{2};\xi _{2}^{2})=Z_{0}(\xi _{2}^{2},\xi
_{1}^{2},p^{2};\xi _{1}^{2})+\ln \left( \frac{\xi _{1}^{2}}{\xi _{2}^{2}}%
\right) .  \label{ref}
\end{equation}
to change the scale of \ the functions. For an arbitrary scale $\eta $ we
will have 
\begin{equation}
Z_{0}(\xi _{1}^{2},\xi _{2}^{2},p^{2};\eta ^{2})=Z_{0}(\xi _{1}^{2},\xi
_{2}^{2},p^{2};\xi _{2}^{2})+\ln \left( \frac{\eta ^{2}}{\xi _{2}^{2}}%
\right) .
\end{equation}

The final result of the IRT manipulation for integral 
\begin{equation}
\int_{\Lambda }\frac{d^{4}k}{(2\pi )^{4}}\frac{1}{(k^{2}-\xi
_{1}^{2})[(p-k)^{2}-\xi _{2}^{2}]}
\end{equation}
can be written as 
\begin{equation}
I_{\log }(\xi _{2}^{2})-\frac{i}{(4\pi )^{2}}Z_{0}(\xi _{1}^{2},\xi
_{2}^{2},p^{2};\xi _{2}^{2})
\end{equation}
or 
\begin{equation}
I_{\log }(\xi _{1}^{2})-\frac{i}{(4\pi )^{2}}Z_{0}(\xi _{2}^{2},\xi
_{1}^{2},p^{2};\xi _{1}^{2})
\end{equation}
or 
\begin{equation}
I_{\log }(\eta ^{2})-\frac{i}{(4\pi )^{2}}Z_{0}(\xi _{2}^{2},\xi
_{1}^{2},p^{2};\eta ^{2})
\end{equation}

Let us now proceed to the explicit verification of the Ward identities (\ref
{iw1}) and (\ref{iw2}) and show that an exchange of mass (scale) in the
finite part (\ref{ref}) precisely corresponds to a mass exchange in the
counterterms (\ref{red}). After all calculations we can write the one loop
Ward identity (\ref{iw1}) as 
\[
-16m^{2}G^{2}[I_{\log }(m^{2})-\frac{i}{(4\pi )^{2}}%
Z_{0}(m^{2},m^{2},p^{2};m^{2})]
\]
\[
+F^{2}\lambda ^{2}\{18[I_{\log }(M^{2})-\frac{i}{(4\pi )^{2}}%
Z_{0}(M^{2},M^{2},p^{2};M^{2})]
\]
\[
+6[I_{\log }(\mu ^{2})-\frac{i}{(4\pi )^{2}}Z_{0}(\mu ^{2},\mu
^{2},p^{2};\mu ^{2})]
\]
\begin{equation}
-4[I_{\log }(M^{2})-\frac{i}{(4\pi )^{2}}Z_{0}(\mu ^{2},M^{2},p^{2};M^{2})]\}
\label{z}
\end{equation}
must be equal to 
\[
-16m^{2}G^{2}[I_{\log }(m^{2})-\frac{i}{(4\pi )^{2}}%
Z_{0}(m^{2},m^{2},p^{2};m^{2})]
\]
\[
+F^{2}\lambda ^{2}\{[14I_{\log }(M^{2})+6I_{\log }(\mu ^{2})]
\]
\begin{equation}
+\frac{i}{(4\pi )^{2}}[4Z_{0}(\mu
^{2},M^{2},p^{2};M^{2})-18Z_{0}(M^{2},M^{2},p^{2};M^{2})-6Z_{0}(\mu ^{2},\mu
^{2},p^{2};\mu ^{2})]\}
\end{equation}
Note that in the last term of eq.(\ref{z}) we have the freedom to write it
as 
\begin{equation}
-4[I_{\log }(\mu ^{2})-\frac{i}{(4\pi )^{2}}Z_{0}(M^{2},\mu ^{2},p^{2};\mu
^{2})]
\end{equation}
One can easily verify that whatever the choice of scale (including an
arbitrary one), in the finite part by the relation (\ref{ref}), the
counterterms are automatically adjusted by the relation (\ref{red}) in such
a way that the Ward identity is always completely satisfied. This becomes
more clear in the Ward Identity (\ref{iw2}), which can be expressed as 
\[
\tau _{a}\gamma _{5}mG^{2}\{3[I_{\log }(m^{2})-\frac{i}{(4\pi )^{2}}%
Z_{0}(\mu ^{2},m^{2},p^{2};m^{2})]
\]
\begin{equation}
-[I_{\log }(m^{2})-\frac{i}{(4\pi )^{2}}Z_{0}(M^{2},m^{2},p^{2};m^{2})]\}
\end{equation}
which must be equal to 
\[
\tau _{a}\gamma _{5}mG^{2}\{[I_{\log }(m^{2})-\frac{i}{(4\pi )^{2}}%
Z_{0}(M^{2},m^{2},p^{2};m^{2})]
\]
\[
+[I_{\log }(m^{2})-\frac{i}{(4\pi )^{2}}Z_{0}(\mu ^{2},m^{2},p^{2};m^{2})]
\]
\begin{equation}
+\frac{i}{(4\pi )^{2}}2[Z_{0}(M^{2},m^{2},p^{2};m^{2})-Z_{0}(\mu
^{2},m^{2},p^{2};m^{2})]\}
\end{equation}
where we have used the fermion mass for commodity only.

\section{Conclusion}

In the calculation it becomes apparent the delicate and essential connection
between divergent and finite parts of amplitudes and the examples show how
to use mass scales identities which are absolutely necessary to manipulate
graphs involving several masses in a way as to show its equivalence to
others involving different (than the previous) masses.

One of the advantages of the present technique is that we have all
counterterms in an explicit form. This simplifies the renormalization
procedure, since this can be done directly in the Lagrangian. The
renormalization procedure due to some symmetry require counterterms with the
same mass and then we can introduce the arbitrary scale lambda, which has
been shown to play the role of the sliding scale of the theory (see Ref.\cite
{irt}).

{\bf Acknowledgments}

The work of \ M.C. Nemes\ was partially supported by CNPq. S.R. Gobira would
like to thank C. M. G.de Sousa and Paulo H. L. Martins for some discussions.

\end{document}